\documentclass[twocolumn,tighten]{aastex631}

\usepackage{xcolor}

\begin{document}
\title{First Calculations of Starspot Spectra based on 3D Radiative Magnetohydrodynamics Simulations}

\author[0000-0003-3490-6532]{H. N. Smitha}
\affiliation{Max-Planck-Institut für
Sonnensystemforschung, Justus-von-Liebig-Weg 3, 37077 Göttingen, Germany}

\author[0000-0002-8842-5403]{Alexander I. Shapiro}
\affiliation{Max-Planck-Institut für
Sonnensystemforschung, Justus-von-Liebig-Weg 3, 37077 Göttingen, Germany}
\affiliation{University of Graz, Institute of Physics, Universitätsplatz 5, 8010 Graz, Austria}

\author[0000-0002-0929-1612]{Veronika Witzke}
\affiliation{Max-Planck-Institut für
Sonnensystemforschung, Justus-von-Liebig-Weg 3, 37077 Göttingen, Germany}
\affiliation{University of Graz, Institute of Physics, Universitätsplatz 5, 8010 Graz, Austria}

\author[0000-0002-6087-3271]{Nadiia M. Kostogryz}
\affiliation{Max-Planck-Institut für
Sonnensystemforschung, Justus-von-Liebig-Weg 3, 37077 Göttingen, Germany}

\author[0000-0001-8217-6998]{Yvonne C. Unruh}
\affiliation{Department of Physics, Imperial College London, London SW7 2AZ, UK}

\author[0000-0002-6568-6942]{Tanayveer S. Bhatia}
\affiliation{Max-Planck-Institut für
Sonnensystemforschung, Justus-von-Liebig-Weg 3, 37077 Göttingen, Germany}

\author[0000-0001-9474-8447]{Robert Cameron}
\affiliation{Max-Planck-Institut für
Sonnensystemforschung, Justus-von-Liebig-Weg 3, 37077 Göttingen, Germany}
\author[0000-0002-6892-6948]{Sara Seager}
\affiliation{Department of Physics and Kavli, 
Institute for Astrophysics and Space Research,
Massachusetts Institute of Technology,Cambridge, MA 02139, USA}
\affiliation{Department of Earth, Atmospheric and Planetary Sciences, Massachusetts
Institute of Technology, Cambridge, MA 02139, USA}
\affiliation{Department of Aeronautics and Astronautics, Massachusetts Institute of
Technology, 77 Massachusetts Avenue, Cambridge, MA 02139, USA}

\author[0000-0002-3418-8449]{Sami K. Solanki}
\affiliation{Max-Planck-Institut für
Sonnensystemforschung,  Justus-von-Liebig-Weg 3, 37077 Göttingen, Germany}

\begin{abstract}
Accurate calculations of starspot spectra are essential for multiple applications in astronomy. The current standard is to represent starspot spectra by spectra of stars that are cooler than the quiet star regions. This implies approximating a starspot as a non-magnetic 1D structure in radiative-convective equilibrium, parametrizing convective energy transport by mixing length theory. It is the inhibition of convection by the starspot magnetic field that is emulated by using a lower spot temperature relative to the quiet stellar regions. Here, we take a different approach avoiding the approximate treatment of convection and instead self-consistently accounting for the interaction between matter, radiation, and the magnetic field. We simulate spots on G2V, K0V, M0V stars with the 3D radiative magnetohydrodynamics code MURaM and calculate spectra ($R \approx 500$ from 250~nm to 6000~nm) using ray-by-ray radiative transfer with the MPS-ATLAS code. We find that the 1D models fail to return accurate umbral and penumbral spectra on K0V and M0V stars where convective and radiative transfer of energy is simultaneously important over a broad range of atmospheric heights rendering mixing length theory inaccurate. However, 1D models work well for G2V stars, where both radiation and convection significantly contribute to energy transfer only in a narrow region near the stellar surface.  Quantitatively, the 1D approximation leads to errors longward of 500 nm of about 50\% for both umbral and penumbral flux contrast relative to quiet star regions on M0V stars, and less than  2\% (for umbrae) and 10\% (for penumbrae) for G2V stars.  
\end{abstract}


\section{Introduction} 
\label{sec:intro}
The action of a stellar dynamo \citep{charbonneau2020} leads to the formation of magnetic field which emerges to stellar surfaces in a form usually described by flux tubes \citep{Solanki2006}. The magnetic field modifies thermodynamic and dynamic properties of stellar atmospheres and leads to the formation of active regions. They consist of facular regions which are formed by ensembles of small magnetic flux tubes and often also contain spots which are formed by bigger flux tubes \citep[see, e.g. detailed discussion in][]{Solanki2013}. The radiative spectra emitted by faculae and spots are different from those emitted by quieter stellar regions. 
This leads to a variety of phenomena. One of the most exciting among them is stellar brightness variability caused by changes in coverage of the visible stellar disk  by magnetic features \citep[e.g. due to the activity cycles or due to stellar rotation, see][]{Basri2021}. The observations of stellar brightness variability revealed a number of important stellar properties, e.g. it allowed determination of stellar rotation periods \citep{McQuillan2014, Santos2019, Reinhold2023} and properties of magnetic flux emergence on stellar surfaces \citep{Schrijver2020, Nemec2022}.

Recently it has also become clear that magnetic features, while helping us to study stars, significantly complicate the detection and characterisation of exoplanets and their atmospheres. For example, the main method for studying  exoplanetary atmospheres currently is the measurement of its transmission spectrum, i.e. apparent dependence of the planetary radius on the wavelength \citep{Seager2000}. The planet  appears bigger at the wavelengths where its atmosphere actively absorbs or scatters stellar radiation so that the transmission spectra allow us to determine the composition of planetary atmospheres. Recent studies, however, revealed that planetary signal can be overwhelmed by contamination from magnetic features unocculted during the transit \citep[see, e.g.,][]{SAG21}. In particular, it is currently impossible to distinguish water in atmospheres of exoplanets and in spots on the host star \citep{Barclay2021, Moran2023}.

The community has developed a number of approaches to circumvent the magnetic contamination of transmission spectra \citep{SAG21, TRAPPIST_rev}. Regardless of the specific details, all these approaches require precise knowledge of both facular and spot spectra. Since their spectra cannot yet be readily observed, the only way to obtain them is to rely on models. 
In the absence of realistic 3D radiative magnetohydrodynamics (MHD) simulations of faculae and spots, the community had to rely on 1D radiative-equilibrium models with a parameterised treatment of convection,  approximating faculae by warmer non-magnetic stars and spots by cooler non-magnetic stars \citep{SAG21}. These models, however, do not account for the complex interaction between magnetic field, plasma, and radiation. \cite{Witzke2022} and \cite{Norris2023} recently showed that 1D radiative-equilibrium models dramatically fail to reproduce the spectral profile of the facular contrast revealed by comprehensive 3D radiative MHD simulations.

In this paper, we for the first time calculate spot spectra based on the MHD simulations of G2V, K0V, and M0V stars. Namely, we use 3D MHD simulations of spots produced with the 3D radiative MHD code MURaM \citep{Voegler2005} and compute their spectra using the radiative transfer code MPS-ATLAS \citep{Witzke2021}. Based on the results of these calculations we also assess the applicability of 1D models employed until now to compute starspot spectra. The current paper is limited to these three spectral types. Later-type M-dwarfs will be addressed in a separate paper since their simulations require an update of molecular opacities in MURaM and MPS-ATLAS.

\begin{figure*}
    \centering
    \includegraphics[width=\textwidth]{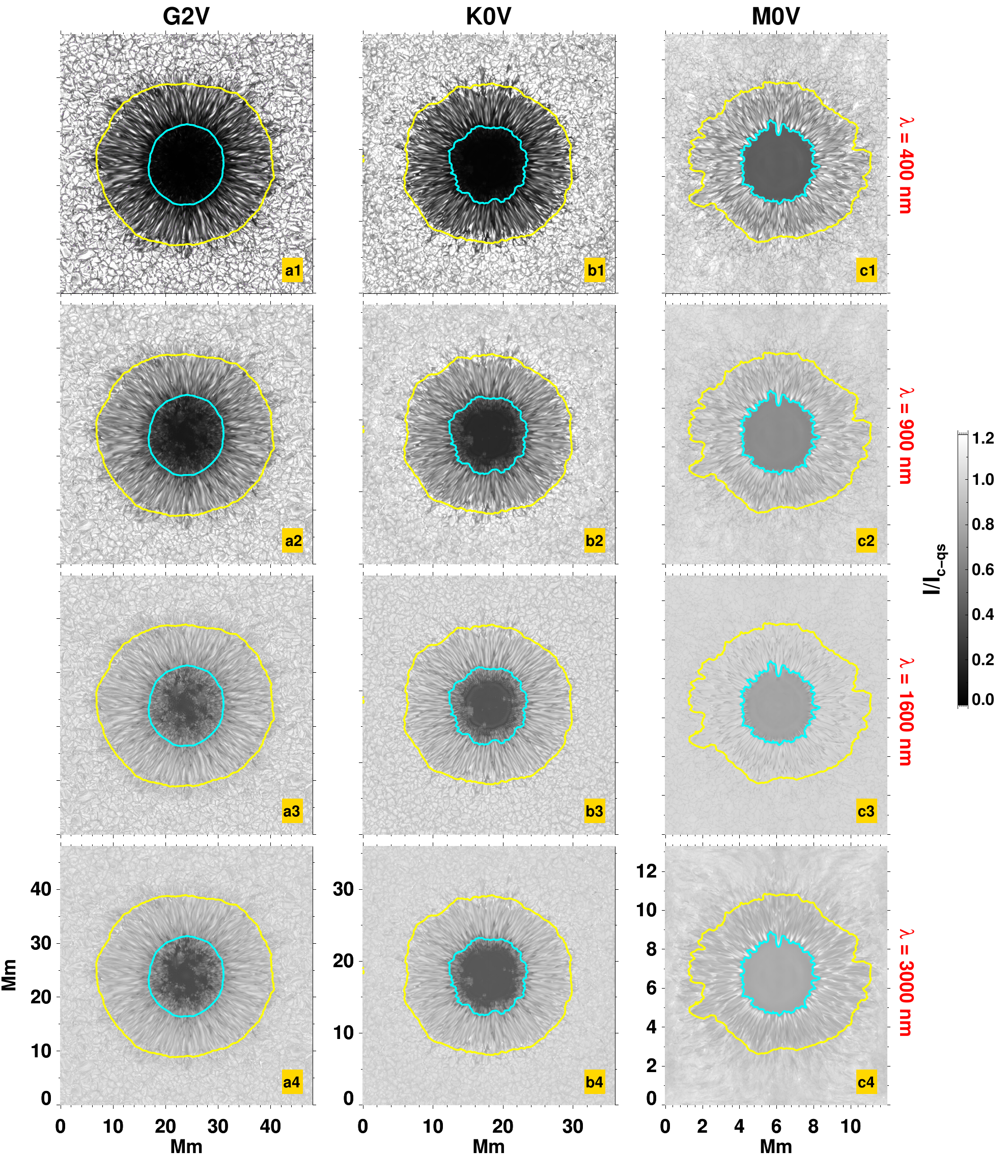}
    \caption{Disk center {intensity} images of the spots for G2, K0 and M0 dwarfs at four representative wavelengths computed with the MPS-ATLAS code. The yellow and cyan contours indicate the masks that were used to select the penumbral and umbral regions in each case. All images have been normalized to their respective spatially averaged quiet region intensities. The gray scale indicates relative intensities between 0 (black) and 1.2 (white) as indicated by the color bar. }
    \label{fig:fig1}
\end{figure*}

\section{Method}
\label{sec:method}

Our calculations are performed in two steps. In the first step (see Sect.~\ref{subsect:MHD}), the atmospheric structures of spots on \mbox{G2-,} K0-, and M0-dwarfs are simulated with the radiative 3D MHD MURaM code. In the second step (see Sect.~\ref{subsection:RTray}), we use the MPS-ATLAS code to calculate spectra from these atmospheres. 

\subsection{{MHD simulations}}
\label{subsect:MHD}

In this section, we outline some of the key details about the MHD simulations used in our work and then in the next section continue with the description of the spectral calculations with the MPS-ATLAS code. A detailed description of MURaM simulations of spots is given in \citet[][in prep.]{Bhatia2024}.

The MURaM simulations we use here are an extension of the work by \cite{Panja2020} who performed simulations of spots in a slab geometry. Here we use a principally similar setup for the magnetic field except that our flux tubes are cylindrical, {as described in \citet[][in prep.]{Bhatia2024}}.
To form an adequate penumbra, we have followed the approach of \cite{Rempel2012}. Namely, we started the simulations in a large box with two spots of opposite polarities using a coarse grid both in horizontal and vertical directions.  After the spots stabilized and formed penumbrae we isolated one of the spots and continued running it at doubled resolution to better simulate the structure of the spot. The final dimensions and resolutions for each of the spectral types are given in Table~\ref{tab:table1}. At the top
boundary we made the magnetic field more horizontal than a
potential field by a factor of 3. This allowed us to preserve a decent sized penumbra. 

\begin{deluxetable*}{cccccccccccc}
\tablecaption{Details of the simulation boxes for the G2V, K0V and M0V stars. \label{tab:table1}}
\tabletypesize{\scriptsize}
\tablenum{1}
\tablehead{ & \multicolumn{3}{c}{Dimensions (Mm)} & \multicolumn{3}{c}{Grid spacing (km)} & & \multicolumn{3}{c}{$T_{\rm eff}$ (K)} & \\
\tableline
\colhead{Spectral type} & 
\colhead{$x$}&
\colhead{$y$}& 
\colhead{$z$}&
\colhead{$\Delta x$} &
\colhead{$\Delta y$} &
\colhead{$\Delta z$} & 
\colhead{No. of cubes} & 
\colhead{{Quiet region}} &
\colhead{{Penumbra}} &
\colhead{{Umbra}} &
\colhead{$\log {\rm g}$}}
         \startdata
         G2V & 48 & 48  & 6 & 48  & 48 & 20 & 9 & 5810 & 5102 & 3819 & 4.438\\
         K0V & 36 & 36 & 3  & 36  & 36 & 10 & 9 & 4965 & 4410&3401& 4.609\\
         M0V & 12 & 13.3& 1.2 & 13.3 & 13.3 & 4 & 8 & 3696 & 3609 &3399 & 4.826  \\
         \enddata
    \end{deluxetable*}
The two-spot simulations as well as initial phase of isolated spot simulations have been performed assuming gray radiative transfer. {Once the isolated spot simulations have stabilized, we switch to a more accurate radiative transfer scheme and use a multi-group approach \citep{Nordlund1982} with four groups adapted from \cite{beeck_2013}. After the non-gray simulations stabilized we saved about ten simulation cubes separated by 10 minutes each (see Table~\ref{tab:table1}). For each of these cubes we computed the emergent spectra between 200\,nm and 6000\,nm with the MPS-ATLAS code using the ray-by-ray approach discussed in Sect.~\ref{subsection:RTray}.}

\subsection{{Spectral synthesis: Ray-by-ray radiative transfer}}
\label{subsection:RTray}

A MURaM simulation provides the atmospheric temperature $T$, density $\rho$,  velocity field $v$ and the magnetic field ${\bf B}$ on a 3-D grid. Values at any location in the domain can then be obtained by interpolation. We calculate spectra  under the assumption of local thermal equilibrium (LTE) which  allows a ray-by-ray solution of the radiative transfer. We perform this solution with the MPS-ATLAS code specifically developed for a fast radiative transfer along the ray \citep{Witzke2021}. 
{The spectrum at any disk position is calculated by solving the radiative transfer along appropriately inclined rays 
 and then considering the spatial average of the resulting intensities.}
For example the spectrum at disk centre involves rays which are oriented along the vertical axis. The rays will be inclined with respect to the vertical when we look away from disk centre. This is achieved in practice by rotating the cube by an angle corresponding to the disk position chosen for spectral synthesis before interpolating to the vertical rays. 

Such a ray-by-ray solution is often refereed in the literature as 1.5D approach \citep[see for example,][]{Holzreuter2015,Smitha2017, Smitha2021} since it does not account for the cross-talk between different rays. This cross-talk, however, only becomes essential either when LTE approximation breaks or when the atmospheric structure has to be computed simultaneously with solving the radiative transfer (see, Sect.~\ref{subsection:RT1D}). We reiterate that all essential atmospheric properties are pre-computed with the MURaM code. Consequently, the ray-by-ray approach enables a full formal solution of the radiative transfer, capturing all the 3D complexities introduced by the dynamic nature of MURaM simulations. The emergent spectra from the MHD cubes presented in Figures~\ref{fig:fig1} - \ref{fig:fig6} and discussed through out the paper are computed using this approach.

All in all, we derive emergent intensities at ten different limb distances from  $\mu=1.0$ (disc centre) to $\mu=0.1$ in steps of $\Delta\mu=0.1$ ($\mu = \cos\theta$, where $\theta$ is the angle between the line of sight and the stellar surface normal). Since molecular spectral lines become increasingly important at low temperatures we consider the following molecules and their most important isotopes in the MPS-ATLAS code calculations: $\rm H_2$, CH, NH, $\rm C_2$, CN, CO, OH, MgH, SiH, SiO, TiO, and $\rm H_2O$. Furthermore, for the continuous opacity, we takes the following contributors into account: Free-free (ff) and bound-free (bf) transitions in $\rm H-$, H2, He, $\rm He-$, C, N, O, Ne, Mg, Al, Si, Ca, Fe, the molecules CH, OH and NH, and their ions.

\subsection{{Spectral synthesis: 1D non-magnetic radiative convective equilibrium model}}
\label{subsection:RT1D}

In addition to the solution of the radiative transfer along the ray the MPS-ATLAS code is capable of  generating the 1D radiative-convective equilibrium models (hereafter, RE models).
In this case both the atmospheric properties, $T$ and $\rho$, as well as the radiation field are determined together. The problem becomes one of finding an atmosphere which is both in hydrostatic equilibrium and thermal equilibrium (i.e. where the cooling and heating at each point in the atmosphere are in balance). It is traditionally solved by iterating the atmospheric structure \citep[see, e.g.,][for the detailed description of the solution implemented in the family of ATLAS codes]{Kurucz1970}. 

The MPS-ATLAS code treats convective energy transfer using  mixing length approximation \citep[see][for the details of the numerical implementation]{Castelli1997}. We emphasise that mixing length approximation is only used for calculating atmospheric models and it is not utilized in ray-by-ray calculations when all atmospheric properties are taken from MURaM cubes (see Sect.~\ref{subsection:RTray}). A more detailed description of the MPS-ATLAS code is given in \citep{Witzke2021} and in \citep{Kostogryz2022, kostogryz2023} who presented  two grids of 1D RE models produced with the MPS-ATLAS code.

We use MPS-ATLAS code for producing 1D models in Section~\ref{sec:1D} where we compare the spectra obtained with these models to those produced from the MURaM cubes and ray-by-ray solution of the radiative transfer.

\section{Spot images}\label{sect:images}

The disk center intensity images of the spots for the three spectral types at four  representative wavelengths are shown in Figure~\ref{fig:fig1}. All images are normalized to their respective mean quiet region (i.e. regions outside of spots) intensities. In every case the quiet region profiles were taken from one hundred rows of pixels at the top and bottom of the intensity image. The yellow and cyan contours represent the masks that were used to select the penumbral and umbral regions. For all the considered cases these masks have been created using reference images at 400 nm. For the G2, K0 and M0 stars the thresholds for the penumbral masks were set at  $0.60, 0.85, 0.88$ times the respective $I_{\rm qs}$. For the umbral masks, the corresponding thresholds were set at $0.15, 0.35, 0.39$.   The masks are fixed for a given spectral type and do not vary with $\mu$ or wavelength. The chosen penumbral and umbral thresholds appear quite low since they are created at a shorter wavelength of 400nm. At longer wavelengths, the same thresholds will correspond to higher values.

A striking feature evident from Figure~\ref{fig:fig1} is the decrease of spot contrasts from G2- to M0-dwarfs. The main reason for the difference in contrast is that although convection in spots is suppressed in all three spectral classes, the spots on M0-dwarfs can still be effectively heated by the radiation {from the deeper layers below}, which transports considerable energy in the M0 star already below the stellar surface. The same is not true for the G2 dwarfs where radiative transfer of energy in spots is not efficient. For a more detailed discussion see \cite{Panja2020}. The decrease of granular size from G2- to M0-dwarfs seen in Figure~\ref{fig:fig1} was earlier reported and discussed by, e.g., \cite{Freytag2012, beeck_2013, Magic2013}.

Figure~\ref{fig:fig1} shows that the visibility of spots (and granulation pattern) also strongly depends on the spectral region of the observations.
Some interesting features include the visibility of umbral sub-structures at $\lambda=1600$\,nm and 3000\,nm in both, G and K dwarfs, and the appearance of fibril-like structures surrounding the spot at $\lambda=3000$\,nm in the M-dwarf.  

\section{Spot spectra} 
\label{sec:spectra}
\begin{figure*}
    \centering
    \includegraphics[width=\textwidth]{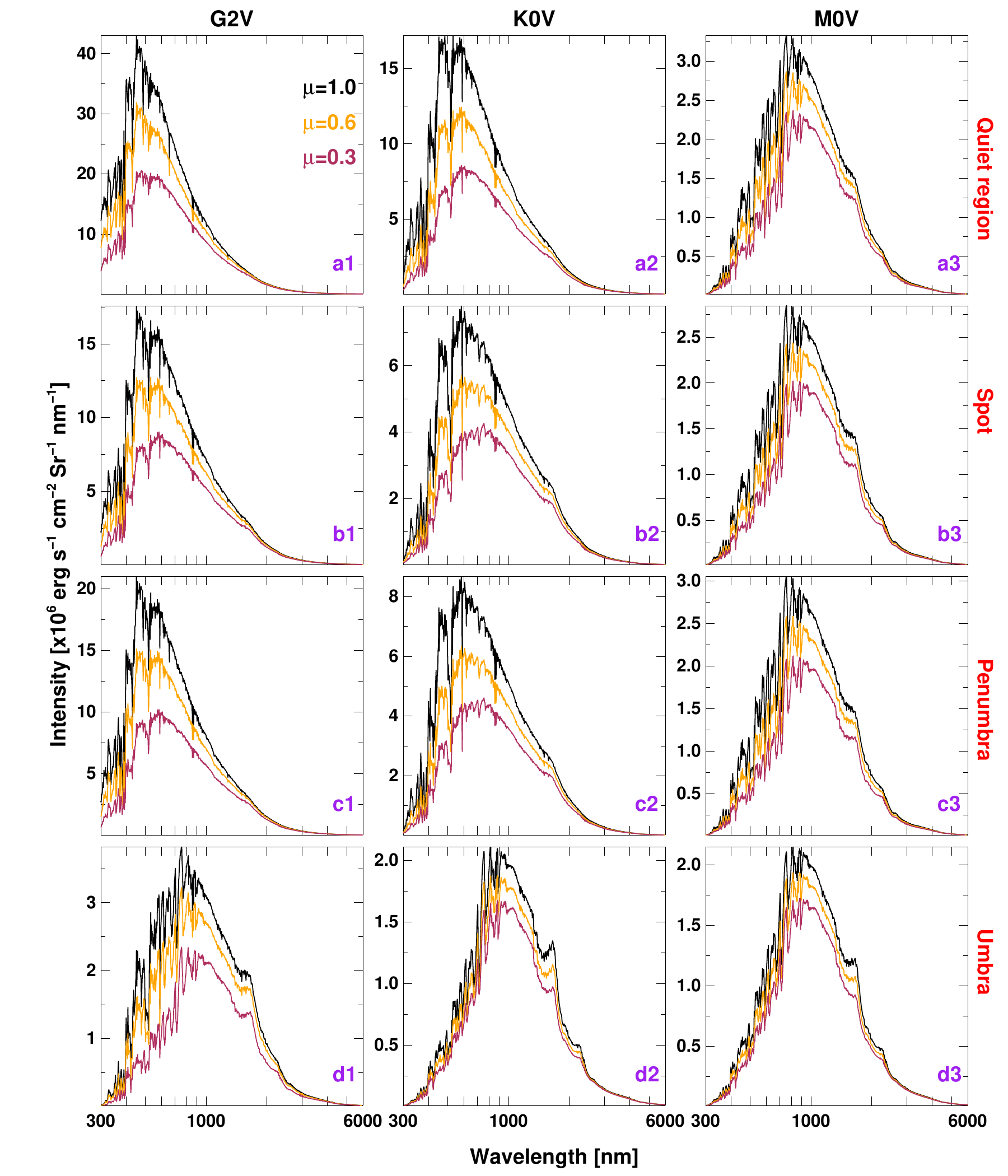}
    \caption{Center-to-limb variation of the spatially averaged intensities from the quiet region, spot, penumbra and umbra for the G2, K0 and M0 dwarfs. The penumbral and umbral regions were selected using the contours shown in Figure~\ref{fig:fig1}. The spot spectra were computed by spatially averaging over the combined penumbral and umbral regions. Similarly, the quiet region spectra were taken from two narrow horizontal bands at the top and bottom of the {intensity images}, away from the spot.}
    \label{fig:fig2}
\end{figure*}

The spatially averaged spectra corresponding to the quiet region, penumbra, umbra and spot (i.e. combined umbra and penumbra) for the G2, K0, and M0 dwarfs are plotted in Figure~\ref{fig:fig2} at three disk positions. The spot spectra are computed by averaging all pixels enclosed within the yellow contour in Figure~\ref{fig:fig1}. The penumbral spectra are from the pixels between the yellow and cyan contours,while the umbral spectra come from the pixels within the cyan contours. 

The large variations in the spectra from $\mu=1.0$ to $\mu=0.3$ are clearly seen not only in the quiet region but also in spots, penumbra and umbra. Out of the three spectral types considered, the M0 dwarf shows relatively weaker dependence on the disk position  (panels a3 - d3 in Figure~\ref{fig:fig2}) due to shallower temperature gradients in their photospheres. The same is true for umbrae, for which the intensity displays a smaller CLV for all three spectral types, but increasingly so for the later spectral types. A detailed analysis on the {center-to-limb variation (CLV)} of the individual spectra and how they compare with the RE models will be presented in a separate paper (Kostogryz et al. in prep).

\begin{figure*}
    \centering
    \includegraphics[width=\textwidth]{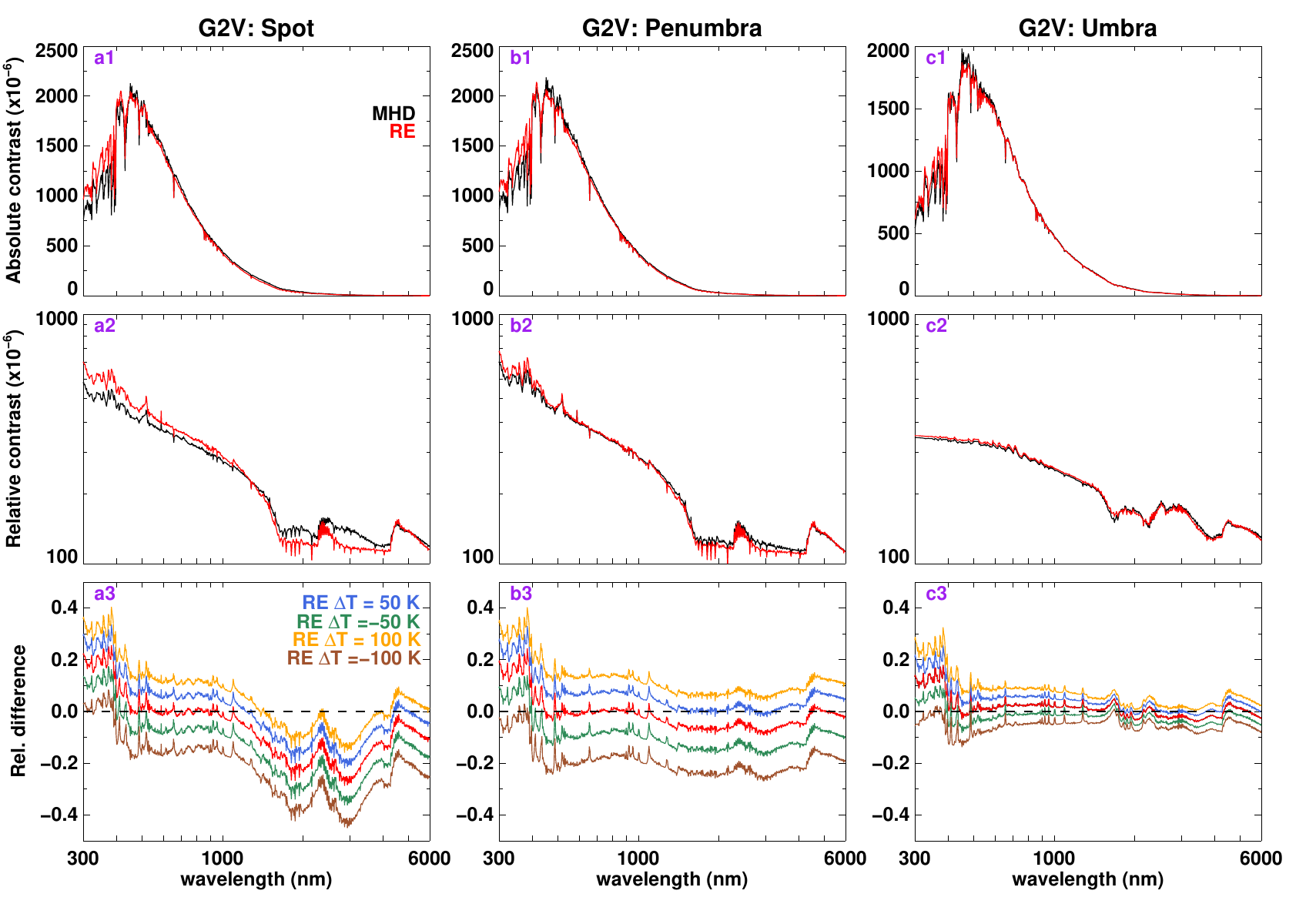}
    \caption{Spot, penumbral and umbral contrasts for the G2 dwarfs. The curves in the top row are the absolute contrasts computed, using Equation~\ref{Eqn:abs_con}, from the 3D MHD simulations, and using radiative equilibrium (RE) models with the same $T_{\rm eff}$ as the MHD case ($T_{\rm eff}^{\rm MHD} - T_{\rm eff}^{\rm RE} = \Delta T_{\rm eff}=0$, red lines). The relative contrasts, computed using Equation~\ref{eq:rel_con}, for the respective cases are plotted on a logarithmic scale in the middle row. To simplify the comparison of the wavelength dependencies returned by MHD and 1D RE calculations we normalize both relative and absolute contrasts by their integrals over the  wavelength domain we show. In other words, the areas under all shown curves for absolute and relative contrasts are equal to unity. In the bottom row are the relative differences, defined in Equation~\ref{eq:rel_diff}, between the MHD contrasts and RE contrasts, shown here for different RE models including those with  $\Delta T_{\rm eff}=\pm 50$\,K, $\pm 100$\,K (see legend). {The relative differences are computed from the absolute contrasts but without the normalization. See Section~\ref{sec:abs_rel_con} for more details.}}
    \label{fig:fig3}
\end{figure*}

\begin{figure*}
    \centering
    \includegraphics[width=\textwidth]{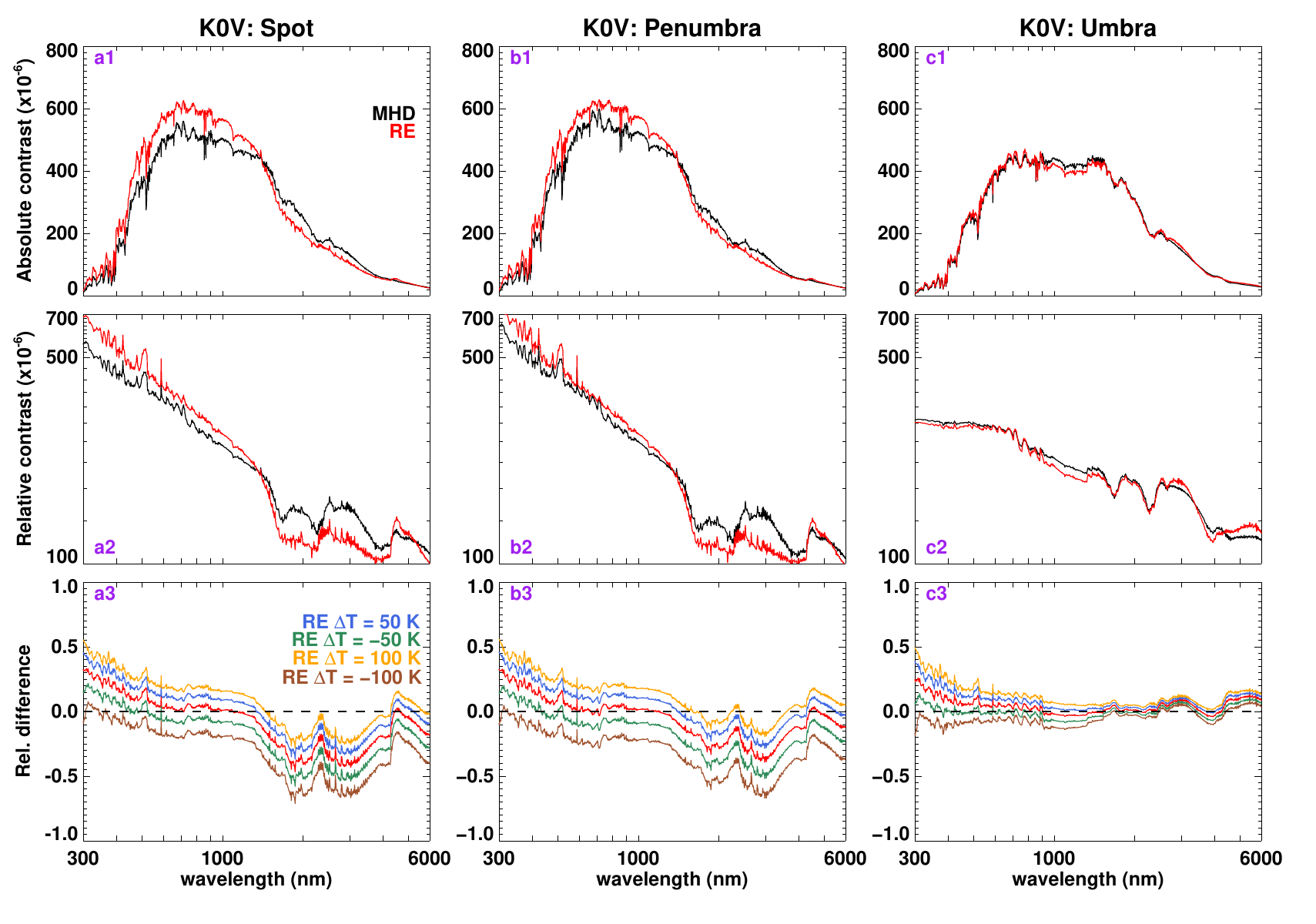}
    \caption{Same as Figure~\ref{fig:fig3} but for K dwarfs.} 
    \label{fig:fig4}
\end{figure*}

\section{Spot contrasts and comparison to 1D radiative equilibrium models} 
\label{sec:1D}
The radiative contrasts of spots relative to quiet stellar regions play a crucial role in a large number of applications, e.g. mitigation of stellar contamination of transmission spectra \citep{Pinhas2018, Barclayetal2021, SAG21, kirketal2024, Damianoetal2024, singetal2024} and stellar variability \citep{Lanza2016, Basri2020, Johnson2021, Nemec2023}. 
Since 3D MHD simulations of spots and their spectra have not been available so far, the contrasts of spots for these applications have been computed using different approximations and simplifications. For example, spots have been routinely represented by non-magnetic models with smaller effective temperatures computed using 1D RE models. 

Here for the first time, we present spot contrasts based on 3D MHD simulations and compare them to the contrasts returned by the RE models. To do this comparison we define effective temperatures of quiet stellar regions, umbra, penumbra, and combined spot model using their disk-integrated spectra (computed based on MURaM simulations). Then we compute RE models with the corresponding temperatures using the MPS-ATLAS code \citep{Witzke2021, kostogryz2023}. These 1D RE models are then used to compute spectral contrasts for all spot components, which are then compared to results based on the MURaM simulations. 

\subsection{Absolute and relative contrasts}
\label{sec:abs_rel_con}
\begin{figure*}
    \centering
    \includegraphics[width=\textwidth]{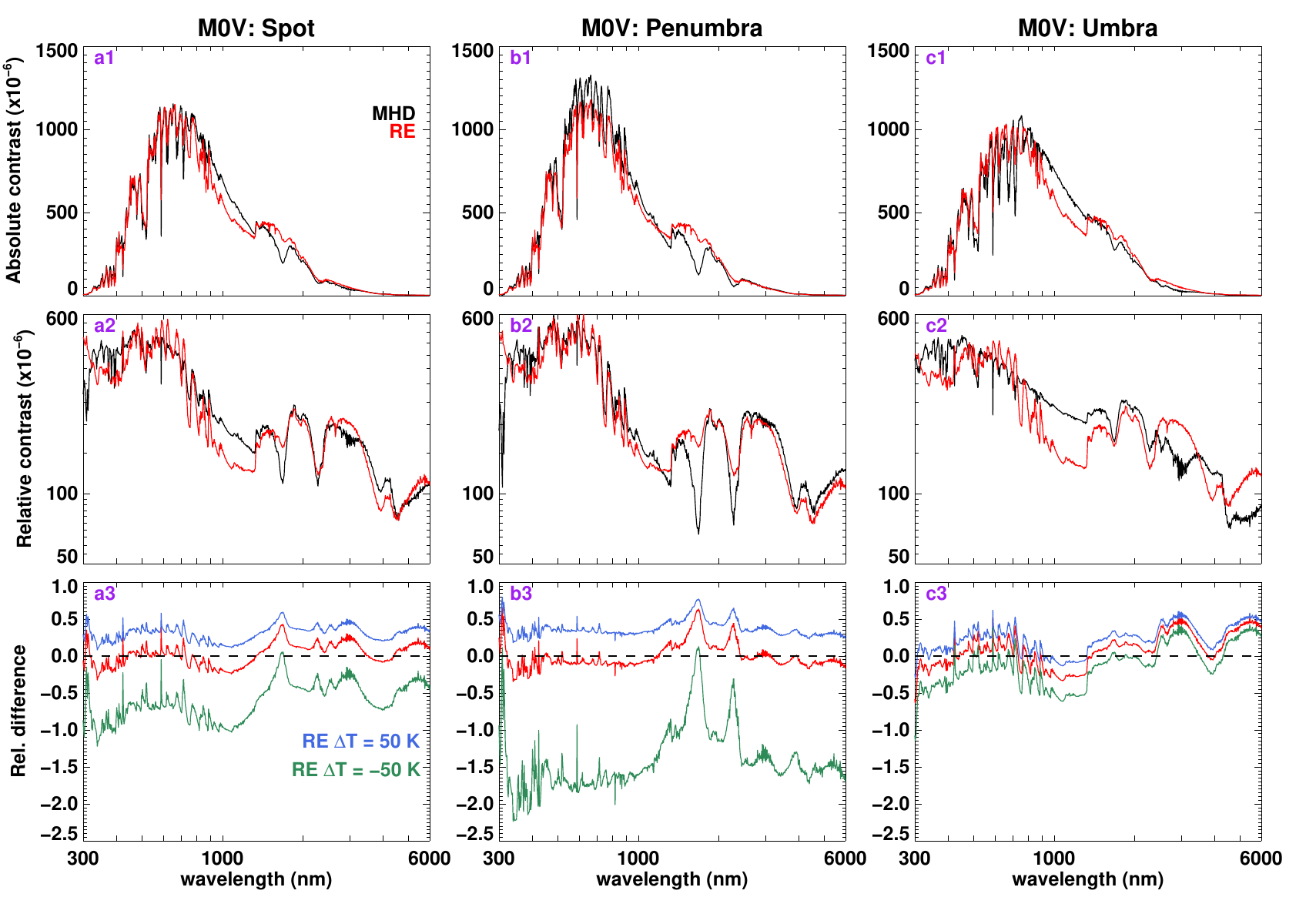}
    \caption{Same as Figure~\ref{fig:fig3} but for M dwarfs. Unlike for the G and K dwarfs, the difference in $T_{\rm eff}$ between the quiet region and the penumbra in the MHD simulations is less than 100\,K. Hence we present a comparison with RE models only for $\Delta T_{\rm eff}=\pm50$\,K.} 
    \label{fig:fig5}
\end{figure*}

\ In Figures~\ref{fig:fig3}, \ref{fig:fig4}, and \ref{fig:fig5}, we present the spot contrasts for the G2, K0 and M0 dwarfs computed based on 3D MHD MURaM simulations (black curves) and from 1D RE MPS-ATLAS models (red curves). Here we define two kinds of contrasts for the magnetic features (umbra, penumbra, spot) relative to the quiet regions (QR). The absolute contrast is defined as:
\begin{equation}
\rm{Absolute\ contrast} = \frac{F_{feat.}(\lambda) - F_{QR}(\lambda)}{\int [F_{feat.}(\lambda) - F_{QR}(\lambda)] d\lambda},
\label{Eqn:abs_con}
\end{equation}
where $\rm F_{feat.}(\lambda), F_{QR}(\lambda),$ are the disk-integrated fluxes of a magnetic feature and quiet region respectively. The denominator in Equation~\ref{Eqn:abs_con} normalizes the absolute contrast to unity over the wavelength range considered here, that is from 300 nm to 6000 nm.  The top panels in Figures~\ref{fig:fig3}--\ref{fig:fig5} show the absolute contrasts for different magnetic features in ppm.

Similarly, we define the relative contrasts as 
\begin{equation}
\rm {Relative \ contrast} = \frac{[F_{feat.}(\lambda) - F_{QR}(\lambda)]/F_{QR}(\lambda)}{\int [F_{feat.}(\lambda) - F_{QR}(\lambda)]/F_{QR}(\lambda)\,d\lambda}.
\label{eq:rel_con}
\end{equation}
The symbols carry the same meaning as in Equation~\ref{Eqn:abs_con}. The relative contrasts for the G2, M0 and K0 stars are shown in the middle panels of Figures~\ref{fig:fig3}--\ref{fig:fig5}. 

\begin{figure*}
    \centering
    \includegraphics[width=\textwidth]{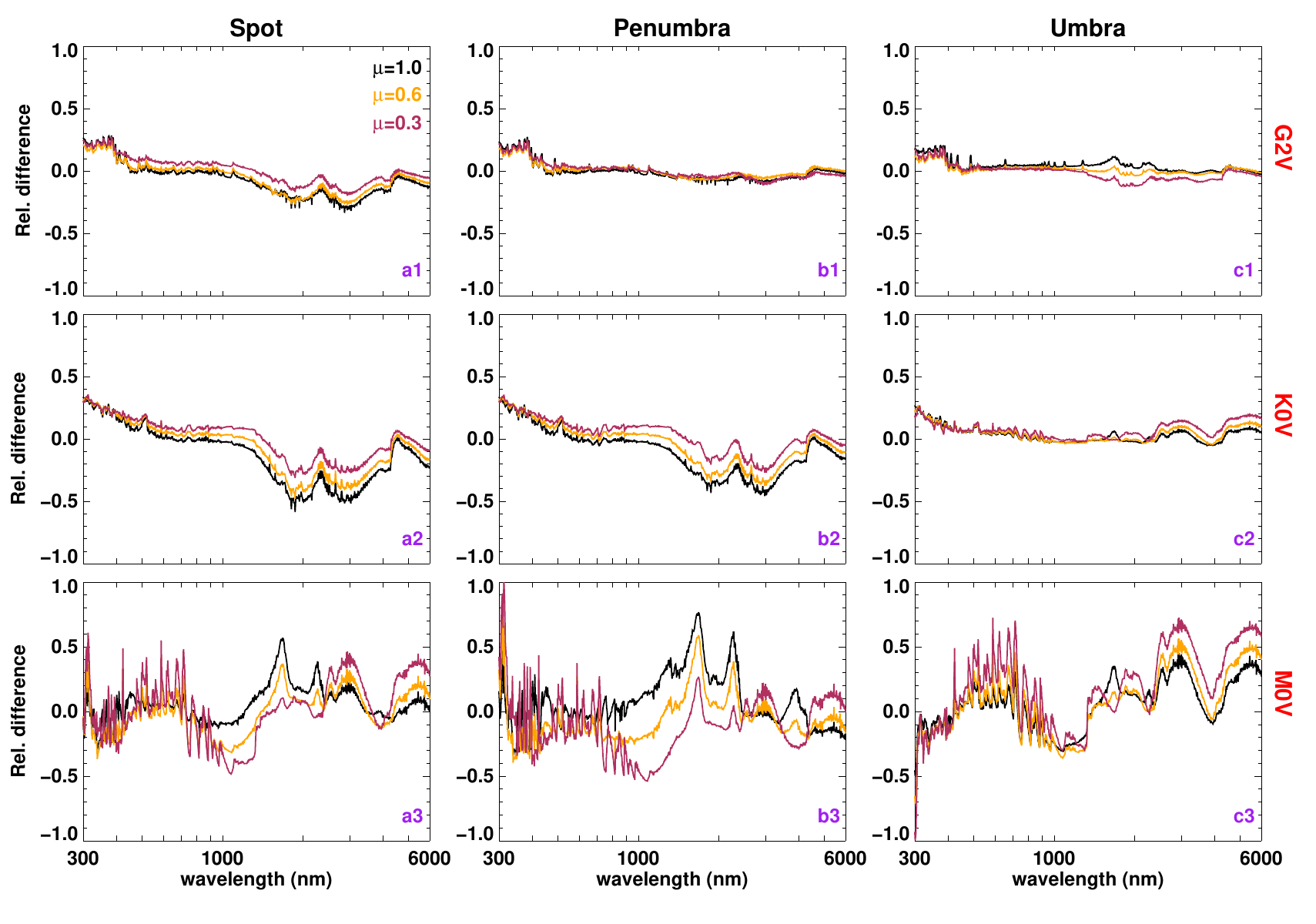}
    \caption{Relative difference between RE and MHD contrasts {(see Equation~\ref{eq:rel_diff})} for the spot, penumbra and umbra for the three spectral types (given on the right of the figure) and three values of the cosine of the heliocentric angle (i.e. angle between the line-of-sight and the local stellar surface normal) $\mu$: 1 (disk centre{, black curve}), 0.6 (intermediate disk position, orange curves), 0.3 (near the limb, maroon curves).}
    \label{fig:fig6}
\end{figure*}

The absolute value of the spot contrast is not essential for many applications. Indeed, the surface coverage of spots is usually unknown, e.g. in transmission spectroscopy applications it is treated as a free parameter.
Thus, the overall multiplication of spot contrast by a wavelength-independent factor can be compensated by the corresponding decrease of the spot surface coverage. 
Thus, to focus on the wavelength dependence of the spot contrast in Figures~\ref{fig:fig3}--\ref{fig:fig5}, both relative and absolute contrasts are normalized such that the area under the curves are equal to unity. 

Finally, in the bottom panels of figures~\ref{fig:fig3}--\ref{fig:fig5}
we plot the relative difference between 1D RE and 3D MHD contrasts defined as:
\begin{equation}
    \rm Relative\ differences = \frac{Contrast_{RE} - Contrast_{MHD}}{Contrast_{RE}}.
    \label{eq:rel_diff}
\end{equation}
 For the computation of relative differences, we use the absolute contrasts defined in Equation~\ref{Eqn:abs_con} but without the normalization factor in the denominator.

For G2V stars the contrasts of umbra and penumbra calculated with 1D RE modelling show close agreement with contrasts calculated using 3D MHD simulations (see middle and right panels of Figure~\ref{fig:fig3}).  This explains why models of solar variability based on 1D RE models of spots, e.g. SATIRE model \citep{Krivova2003, Krivova2008} successfully reproduce the available measurements of solar variability \citep[][]{Solanki2013, Ermolli2013, Yeo2014}. At the same time, a single RE model fails to reproduce spectral contrast of an entire starspot (i.e. combination of umbra and penumbra). This demonstrates that the use of two-component spot models (one component for the umbra and another for the penumbra) is warranted \citep{wenzler2006}.  Approximating the entire spot by a single component as it is routinely done in the literature \citep[see, e.g.,][and references therein]{SAG21} leads to an intrinsic error in the resulting spectrum. The error introduced by representing a spot by a single model atmosphere cannot be compensated for by adjusting its effective temperature since such an adjustment will produce a mere shift of the entire contrast and, thus, is not capable of reproducing spot contrasts at all wavelengths (see bottom panel of Figures~\ref{fig:fig3}). 

\begin{figure*}
    \centering
    \includegraphics[width=0.9\textwidth]{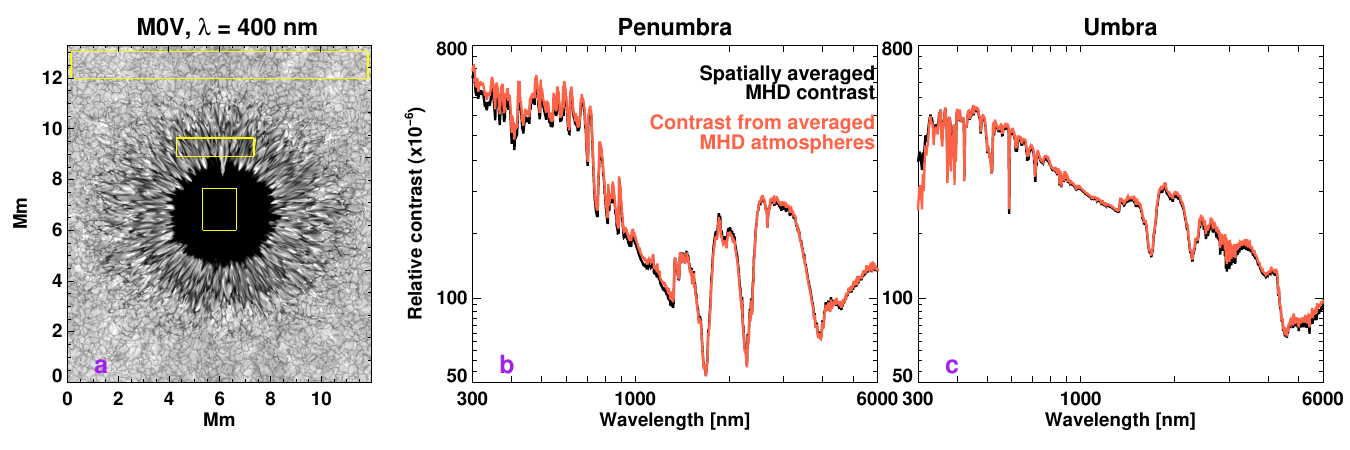}
    \caption{Effects of horizontal inhomogeneities on the contrasts. Comparison between the spatially averaged penumbral and umbral contrasts{(Equation~\ref{eq:rel_con})} from the 3D MHD cube for an M0V-dwarf, and the contrasts computed from the '1D MHD' atmospheres obtained by spatially averaging the atmospheres over the respective yellow boxes in panel a.}
    \label{fig:fig7}
\end{figure*}

In contrast to the G2 case, RE models do not allow an accurate reproduction of umbral and penumbral contrasts for K0 and M0 dwarfs (see, Figures~\ref{fig:fig4}~and~\ref{fig:fig5}, respectively). In particular, the water features seen between 2 and 3 microns in the penumbral contrast of the K0-dwarf are not properly accounted for in RE models. {Also, the RE model fails to recover the spectral slope of the K0 umbral contrast above 4 micron. }
The bottom panel of Figure~\ref{fig:fig4} shows that, similar to the G2 spot case, the noted shortcomings of RE modelling cannot be corrected by adjusting their effective temperatures. For the M0 case, RE modelling dramatically fails for the umbral contrast over the entire spectral domain considered in this study (0.3--6 micron). All in all, 3D MHD modelling is necessary for calculating spectra of spots on K0 and M0 stars.  While the discussion above was limited to the disk-integrated spectra, Figure~\ref{fig:fig6} shows that the overall shortcomings of the RE modelling are present also at different disk positions. 

\subsection{The causes of the discrepancy}
\label{sec:sec4.1}
In the following, we pinpoint the origin of the difference between spot contrasts calculated with the 1D RE and 3D MHD models. There are two critical differences between these two approaches. First, the 1D RE modeling ignores the presence of horizontal substructures in the quiet and active stellar regions (e.g. granulation in the quiet 
regions, filaments and spines in penumbrae, or umbral dots in umbrae). In contrast, these substructures are resolved and accounted for in the 3D MHD simulations. Second, 1D RE models consider a parameterized treatment of convection \citep[e.g. ATLAS and MPS-ATLAS codes use the mixing length approximation by][]{bohm-vitense1958} and do not directly account for the magnetic field. In contrast, 3D MHD models allow self-consistent treatment of the convection and its interaction with the magnetic field. 

We illustrate the relative importance of both these effects in the case of M0 calculations where the deviations between 1D RE and 3D MHD calculations are the largest. To assess the influence of the horizontal substructures on the spot contrast, we take the 3D MURaM cube and horizontally average the vertical columns corresponding to umbra, penumbra, and quiet regions (see Figure~\ref{fig:fig7}) to construct equivalent `1D MHD' models. These 1D MHD models include a proper treatment of convection and the magnetic field, but average out the variations in substructures. {Figure~\ref{fig:fig7} compares the umbral and penumbral contrasts calculated from the 1D MHD models to the 3D MHD contrast computed with a ray-by-ray approach described in Sect.~\ref{sec:method}.} 
Interestingly, both calculations return very similar results (see middle and right panels of Figure~\ref{fig:fig7}), with the differences being much smaller than those seen between 1D and 3D models in \ref{fig:fig5}. This implies that the effect of the horizontal temperature inhomogeneities caused by substructures on the spectra of spots is rather small. 

\begin{figure*}
    \centering
    \includegraphics[width=\textwidth]{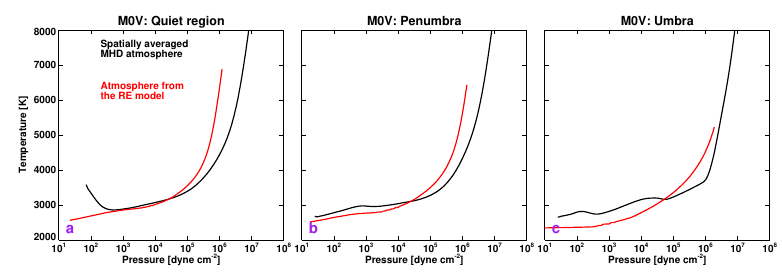}
    \caption{Differences in vertical structure between the spatially averaged one-dimensional MHD atmosphere and the RE models.}
    \label{fig:fig8}
\end{figure*}

In contrast, the comparison of the  `1D MHD' and 1D RE models shown in Figure~\ref{fig:fig8} reveals that these models have very different vertical stratification. In particular, 1D RE modelling overestimates vertical temperature gradients. We attribute it to the failure of the mixing length approximation. Indeed, in contrast to the atmospheres of G-dwarfs where the transition between convective and radiative energy transfer is rather abrupt, convection plays an important role in the entire photosphere of M0 dwarfs \citep{beeck_2013, Panja2020}. This causes the mixing length approximation to fail leading to a less efficient transfer of energy (and, consequently, steeper temperature gradients) than in the realistic MHD simulations.  We note that this is the same effect that causes a decrease of spot contrast towards cooler stars (see Figure~\ref{fig:fig1}).  

Since Figure~\ref{fig:fig7} shows that horizontally averaged '1D MHD' models allow reliable calculations of spectra we give these models in Table~\ref{tab:table2} for the quiet region, penumbra, and umbra. We note, however, that we only tested the goodness of '1D MHD' models for calculations with low spectral resolution. There are additional effects important for the synthesis of individual spectral lines which cannot be adequately accounted for in '1D MHD' models, e.g. flows associated with (magneto-)convection or the effects of the magnetic field on the shapes of spectral lines. 

\begin{deluxetable*}{ccccccccc}
\tablecaption{Tabulated `1D MHD' atmospheres representative of the quiet region, penumbra and umbra for an M0-dwarf. \label{tab:table2}}
\tablenum{2}
\tablehead{\multicolumn{3}{c}{Quiet region} & \multicolumn{3}{c}{Penumbra} & \multicolumn{3}{c}{Umbra} \\ 
\tableline
\colhead{Density} & \colhead{Temperature } & \colhead{pressure} & \colhead{Density} & \colhead{Temperature } & \colhead{Pressure } & \colhead{Density} & \colhead{Temperature} & \colhead{Pressure} \\ 
 \colhead{($\rm{g/cm^3}$)} & \colhead{(K)} & \colhead{($\rm{dyne/cm^{2}}$)} & \colhead{($\rm{g/cm^3}$)} & \colhead{(K)} & \colhead{($\rm{dyne/cm^{2}}$)} & \colhead{($\rm{g/cm^3}$)} & \colhead{(K)} & \colhead{($\rm{dyne/cm^{2}}$)} } 

\startdata
 3.344e-10 & 3.579e+03 & 7.119e+01 & 1.586e-10 & 2.686e+03 & 2.627e+01 & 1.482e-10 & 2.649e+03 & 2.538e+01 \\
 3.885e-10 & 3.469e+03 & 8.039e+01 & 1.645e-10 & 2.668e+03 & 2.698e+01 & 1.547e-10 & 2.655e+03 & 2.653e+01 \\
 5.040e-10 & 3.340e+03 & 1.009e+02 & 1.862e-10 & 2.663e+03 & 3.044e+01 & 1.719e-10 & 2.665e+03 & 2.958e+01 \\
 6.176e-10 & 3.240e+03 & 1.204e+02 & 2.293e-10 & 2.681e+03 & 3.768e+01 & 1.941e-10 & 2.683e+03 & 3.358e+01 \\
 7.506e-10 & 3.145e+03 & 1.427e+02 & 2.940e-10 & 2.704e+03 & 4.844e+01 & 2.183e-10 & 2.688e+03 & 3.773e+01 \\
\enddata
\tablenotemark{ }
\tablenotetext{ }{Table 2 is published in its entirety in a  machine-readable format. A portion is shown here for guidance regarding its form and content. The full table can be accessed at the \href{https://doi.org/10.17617/3.HS2EE6}{Max Planck Digital Library repository}}

\end{deluxetable*}



\begin{deluxetable*}{ccccccccc}
\tablecaption{Tabulated `1D MHD' atmospheres representative of the quiet region, penumbra and umbra for an K0-dwarf. \label{tab:table3}}
\tablenum{3}
\tablehead{\multicolumn{3}{c}{Quiet region} & \multicolumn{3}{c}{Penumbra} & \multicolumn{3}{c}{Umbra} \\ 
\tableline
\colhead{Density} & \colhead{Temperature } & \colhead{pressure} & \colhead{Density} & \colhead{Temperature } & \colhead{Pressure } & \colhead{Density} & \colhead{Temperature} & \colhead{Pressure} \\ 
 \colhead{($\rm{g/cm^3}$)} & \colhead{(K)} & \colhead{($\rm{dyne/cm^{2}}$)} & \colhead{($\rm{g/cm^3}$)} & \colhead{(K)} & \colhead{($\rm{dyne/cm^{2}}$)} & \colhead{($\rm{g/cm^3}$)} & \colhead{(K)} & \colhead{($\rm{dyne/cm^{2}}$)} } 

\startdata
  2.202e-09 & 3.585e+03 & 5.073e+02 & 6.574e-11 &  2.582e+03 & 1.197e+01 & 7.343e-12 & 2.292e+03 & 1.183e+00 \\
  2.604e-09 & 3.581e+03 & 6.002e+02 & 6.832e-11 & 2.579e+03 & 1.232e+01 & 7.657e-12 & 2.297e+03 & 1.235e+00\\
  3.325e-09 & 3.570e+03 & 7.642e+02 & 7.793e-11 & 2.589e+03 & 1.398e+01 & 8.600e-12 & 2.311e+03 & 1.390e+00\\
  4.000e-09 & 3.559e+03 & 9.168e+02 & 9.814e-11 & 2.623e+03 & 1.789e+01 & 9.922e-12 & 2.328e+03 & 1.609e+00\\
  4.747e-09 & 3.549e+03 & 1.087e+03 & 1.313e-10 & 2.662e+03 & 2.436e+01 & 1.131e-11 & 2.345e+03 & 1.839e+00 \\
\enddata
\tablenotemark{ }
\tablenotetext{ }{Table 3 is published in its entirety in a machine-readable format. A portion is shown here for guidance regarding its form and content.  The full table can be accessed at the \href{https://doi.org/10.17617/3.HS2EE6}{Max Planck Digital Library repository}}
\end{deluxetable*}

\section{Conclusions} \label{sec:conclusions}
We showed that 1D RE modeling fails to match the 3D MHD spectral contrasts of umbra and penumbra for M0- and K0-dwarfs.  We attribute the discrepancies to the inability of the mixing length approximation to accurately account for convection and its interaction with the magnetic field in atmospheres of stars cooler than the Sun. In such stars  both convection and radiation play an important role in energy transfer over the entire photosphere \citep[][]{beeck_2013, Panja2020}.

One interesting possibility to test our calculations is provided by observations of the spot-crossing events when a transiting planet crosses a spot on its way across the stellar disk. A number of such events have been already observed with broad-band photometric filters, e.g persistent spot crossing events on TOI-3884 \citep{Almenara2022, Libby-Roberts2023}. Spectral observations of such events (e.g. with the Hubble Space Telescope or James Webb Space Telescope)
are capable of uncovering spot spectra and providing additional observational constraints for the MHD simulations.

The spot spectra provided in this paper can be used for calculating contamination of transmission spectra on M0, K0, G2 stars and are available  at the \href{https://doi.org/10.17617/3.HS2EE6}{Max Planck Digital Library repository}.

\begin{acknowledgments}
AIS, VW, and SS acknowledge support form the European Research Council (ERC) under the European Union’s Horizon 2020 research and innovation program (grant no. 101118581).  This work is partially funded by JWST-AR-03593.
This project has received funding from the European Research Council (ERC) under the European Union's Horizon 2020 research and innovation programme (grant agreement No. 101097844 — project WINSUN); YCU acknowledges funding through Science and Technology Facilities Council (STFC) grant  ST/W000989/1.
\end{acknowledgments}

\bibliographystyle{aasjournal}

\appendix
\section{Comparison with PHOENIX spectra}
\label{sec:app1}
\begin{figure*}
    \centering
    \includegraphics[width=\textwidth]{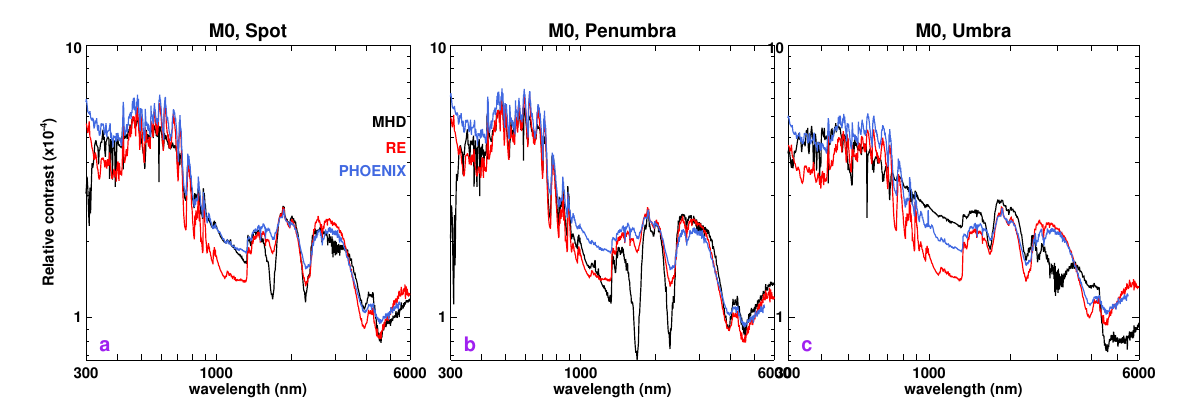}
    \caption{Comparison between relative contrasts computed from the MHD cube, RE models and the Phoenix spectra for the M dwarfs. }
    \label{fig:fig9}
\end{figure*}

In Figure~\ref{fig:fig9} we compare the 3D MHD contrasts for the M0-dwarfs and from the 1D RE models with the contrasts computed from the 1D PHOENIX models \citep[][]{Husser2013}. In the PHOENIX-grid, we take models with effective temperatures closest to our 1D RE models and interpolate them to the exact desired value. The penumbral, umbral and the combined spot contrasts from the PHOENIX models show minor deviations to the 1D RE contrasts \citep[][]{Witzke2021} but overall display a similar trend. Although they lie closer to the spectra from the 3D simulations at some wavelengths, at others they deviate even more than the 1D RE models we considered. Thus the 1D PHOENIX contrasts also show a dramatic failure to match the 3D MHD contrasts on the M0 dwarf. 

\section{Influence of different molecular bands}
\begin{figure*}
    \centering
    \includegraphics[width=\textwidth]{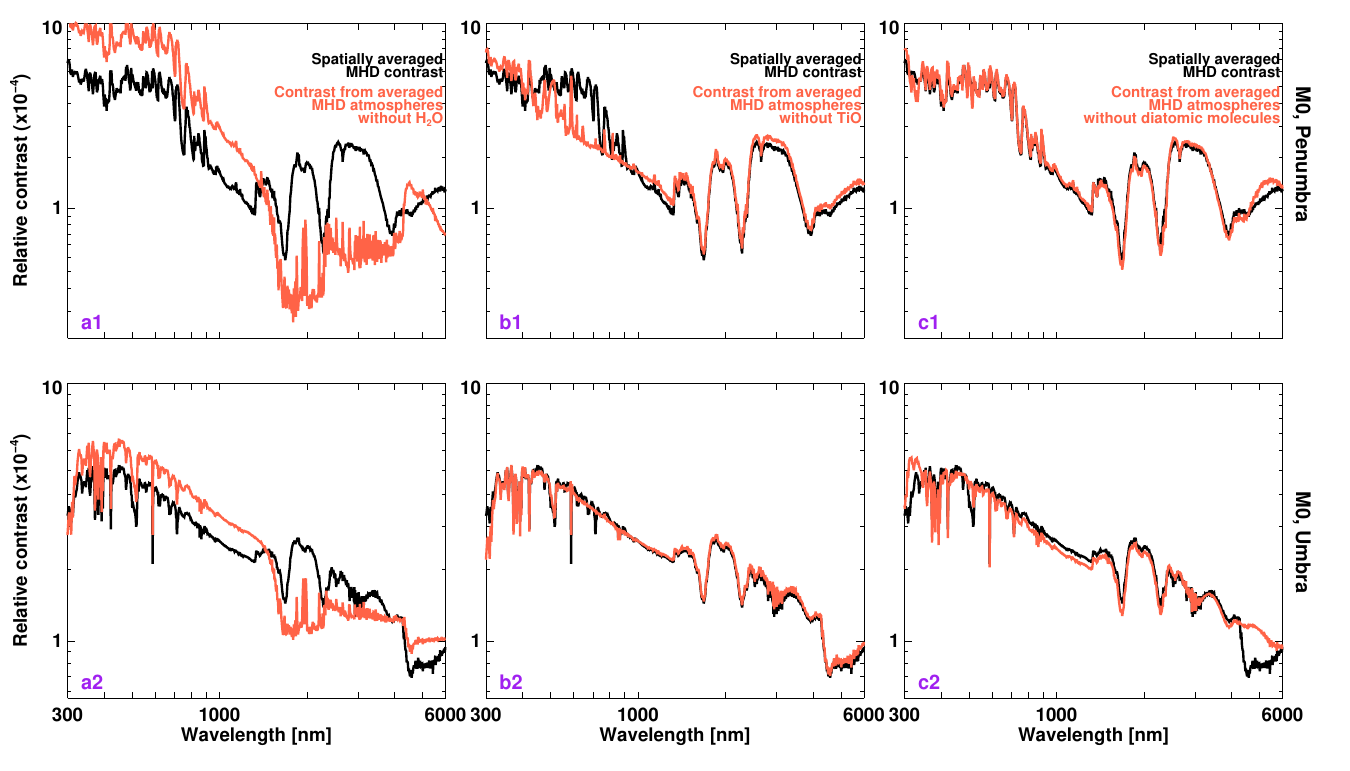}
    \caption{Relative contrasts for the M0 star from the spatially averaged 1D MHD atmosphere computed without the opacities from H$_2$O molecules (panels a1 and a2); without the opacities from the TiO molecules (panels b1 and b2); and without the opacities from the diatomic molecules (panels c1 and c2). The top row shows the penumbral contrasts while the umbral contrasts are plotted in the bottom row.}
    \label{fig:fig10}
\end{figure*}

Since the contrasts from the spatially averaged `1D MHD' closely match the 3D MHD contrasts for the M0-dwarfs (see Section~\ref{sec:sec4.1}, Figure~\ref{fig:fig7}), we now use the `1D MHD' model to investigate the influence of a variety of molecular band on different spectral regions. In Figure~\ref{fig:fig10}, we compute the penumbral and umbral contrasts from the M0-dwarfs by suppressing the opacities from water molecules (panels a1 and a2), TiO molecules (panels b1 and b2) and all other diatomic ($\rm H_2$, CH, NH, $\rm C_2$, CN, CO, OH, MgH, SiH, SiO)  (panels c1 and c2). The absence of water molecules leads to a large deviation in the contrasts over nearly the entire spectral band considered in this study, that is 0.2 - 6.0 microns, in both umbra and penumbra.  While the diatomic molecules mainly affect longward of 4 microns in both umbral and penumbral contrasts, the influence of TiO molecules show up only in penumbral contrasts at both shorter and longer wavelengths.

\section{'1D MHD' model atmospheres for K0-dwarfs} 

\begin{figure*}[h!]
    \centering
    \includegraphics[width=\textwidth]{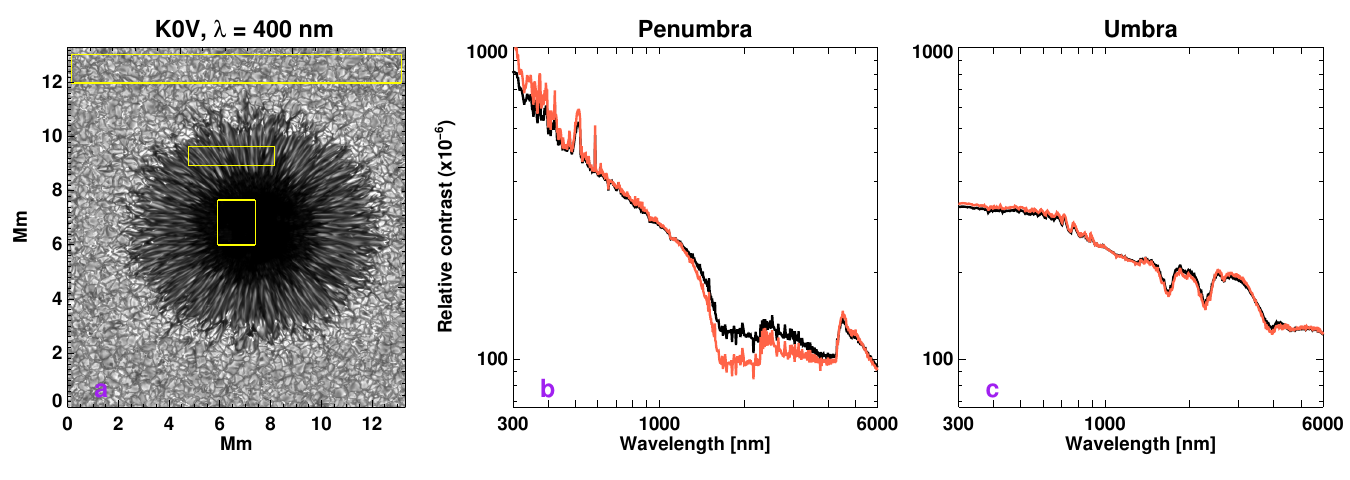}
    \caption{Same as Figure~\ref{fig:fig7} but for a K0-dwarf.} 
    \label{fig:fig12}
\end{figure*}

A similar comparison between the contrasts from the 3D MURaM cube and the '1D MHD' models are presented in Figure~\ref{fig:fig12}. The corresponding '1D MHD' models are given in Table~\ref{tab:table3}.

\section{Variations in spectra between cubes in a time series}
\begin{figure*}
    \centering
    \includegraphics[width=0.45\textwidth]{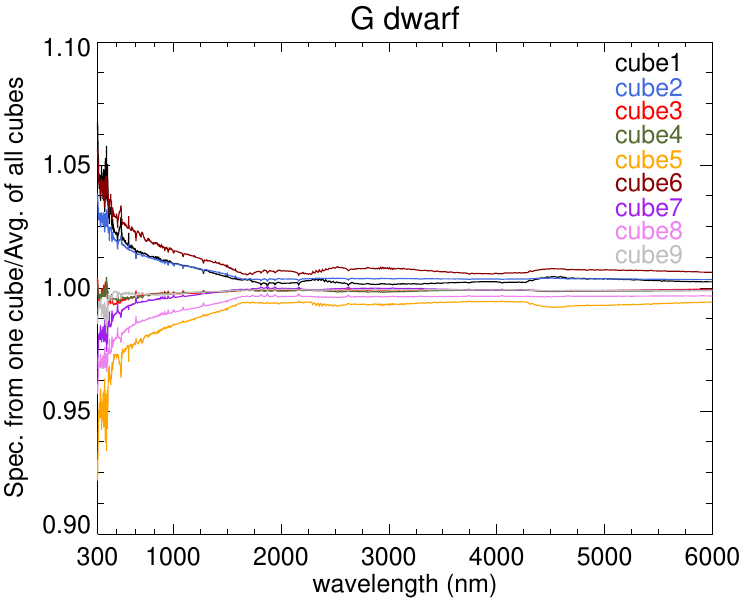}
    \includegraphics[width=0.45\textwidth]{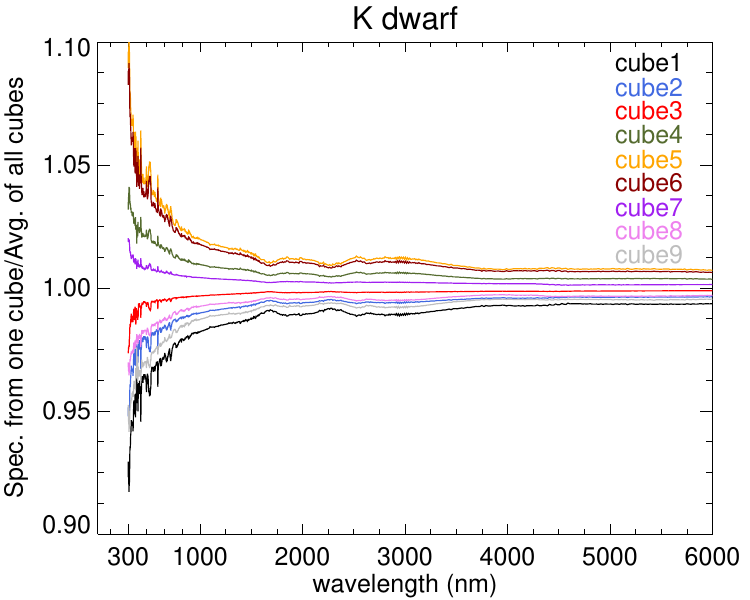}
    \includegraphics[width=0.45\textwidth]{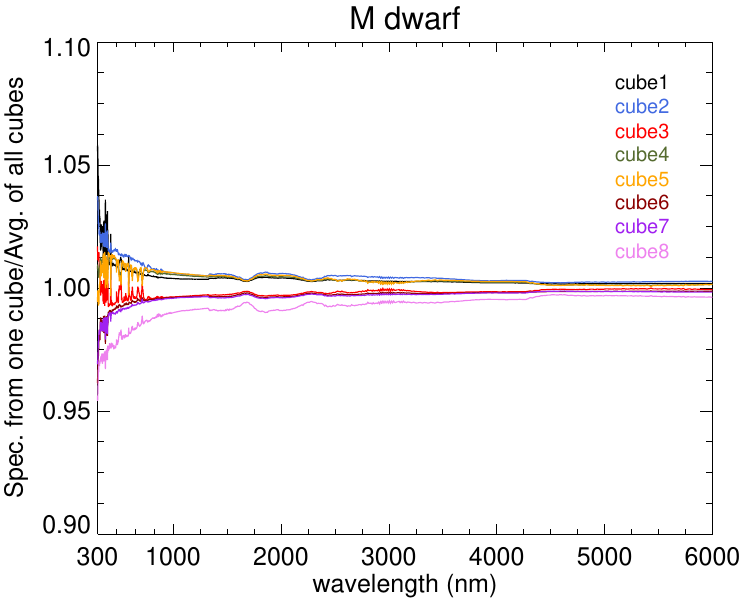}
    \caption{Ratio of spot spectra from every cube used in our analysis to the spectra from the average of all the cubes, shown here for G, K, and M spectral types. } 
    \label{fig:fig11}
\end{figure*}

The 3D MHD contrasts for the G2, M0 and K0-dwarfs discussed in the present paper are all computed using the spectra averaged over a number of cubes from the given time series. Care was taken to select only those cubes whose spectra do not deviate significantly from the mean. The disk-integrated fluxes were then calculated from the average of the selected cubes. The spot spectra from all the different cubes for the G2, K0 and M0-dwarfs are shown in Figure~\ref{fig:fig11}.

\end{document}